# Low-energy Structural Dynamics of Ferroelectric Domain Walls in Hexagonal Rare-earth Manganites

Xiaoyu Wu[1], Urko Petralanda[2], Lu Zheng[1], Yuan Ren[1], Rongwei Hu[3], Sang-Wook Cheong[3*], Sergey Artyukhin[2*], Keji Lai[1*]

1. Department of Physics, University of Texas at Austin, Austin, Texas 78712, USA

2. Quantum Materials Theory, Istituto Italiano di Tecnologia, Genova, Italy

3. Rutgers Center for Emergent Materials and Department of Physics and Astronomy, Rutgers University, Piscataway, NJ 08854

* Correspondence should be addressed to S.-W. C. (sangc@physics.rutgers.edu), S. A. (Sergey.Artyukhin@iit.it) and K. L. (kejilai@physics.utexas.edu)

## Abstract

**Domain walls (DWs) in ferroic materials, across which the order parameter abruptly changes its orientation, can host emergent properties that are absent in the bulk domains. Using a broadband ($10^6$–$10^{10}$ Hz) scanning impedance microscope, we show that the electrical response of the interlocked antiphase boundaries and ferroelectric domain walls in hexagonal rare-earth manganites (h-$R$MnO$_3$) is dominated by the bound-charge oscillation rather than free-carrier conduction at the DWs. As a measure of the rate of energy dissipation, the effective conductivity of DWs on the (001) surfaces of h-$R$MnO$_3$ at GHz frequencies is drastically higher than that at dc, while the effect is absent on surfaces with in-plane polarized domains. First-principles and model calculations indicate that the frequency range and selection rules are consistent with the periodic sliding of the DW around its equilibrium position. This acoustic-wave-like mode, which is associated with the synchronized oscillation of local polarization and apical oxygen atoms, is localized perpendicular to the DW but free to propagate along the DW plane. Our results break the ground to understand structural DW dynamics and exploit new interfacial phenomena for novel devices.**

*Summary sentence: Domain wall dynamics in hexagonal manganites are revealed by impedance microscopy and first-principle calculations.*

# I. Introduction

Domain walls in ferroic materials are the natural interfaces separating domains with different order parameters. Their dynamic responses to external stimuli, which have received tremendous research interest in recent years, play an essential role in determining the material properties. In ferromagnets, for example, the supersonic motion of DWs driven by spin-polarized current leads to the exciting development of magnetic racetrack memories (*1, 2*). The periodic motion of magnetic DWs around the equilibrium position can be induced by a radio-frequency (rf) current, which has been studied to extract the effective DW mass and to enable low-current device operations (*3*). For the electric counterpart, the propagation of ferroelectric DWs under a direct-current (dc) bias has been extensively investigated to understand the switching mechanism (*4-6*). On the other hand, the dynamics of ferroelectric DWs under alternating-current (ac) electric fields have not been thoroughly analyzed. While signatures of ferroelectric DW oscillation, e.g., the dielectric dispersion at the microwave regime (*7-12*), have long been noticed by the scientific community, little is known on the nanoscale dynamics down to the single DW level since conventional bulk measurements inevitably sum up the responses from domains with different polarizations and walls with different orientations (*13, 14*). Spatially resolved studies that address the nanoscale ac response are therefore crucial to explore the underlying physics of such low-energy excitations localized at the DWs, which may be useful for nanoelectronic applications (*15*).

The electrical probing of ferroelectric DW response at the MHz – GHz frequency ($f$) regime may be complicated by the presence of mobile carriers. In order to maintain a spontaneous polarization, the bulk ferroelectric domains are usually highly resistive. Ferroelectric DWs, however, can host anomalous electrical conduction due to the redistribution of carriers, which has indeed been observed by conductive atomic-force microscopy (C-AFM) studies (*16-22*). The difference of dc conductivity between DWs and domains is usually large for charged walls, where free carriers are accumulated or depleted due to the polarization discontinuity (*23, 24*), and small for nominally uncharged walls, where secondary effects such as the flexoelectric coupling (*25*) or the reduction of bandgap (*16*) may take place. In addition, defects such as excess oxygen or oxygen vacancies can also affect the conductivity of DWs and domains through the change of carrier density (*26-29*). Since the contribution from mobile carriers and bound charges at the DWs to the energy loss cannot be separated in a single-*f* measurement, a broadband study is of vital importance to understand the dynamic response of ferroelectric DWs.

In this work, we report the multi-*f* impedance microscopy experiments and theoretical analysis on the DWs of single-crystalline rare-earth hexagonal manganites (h-$R$MnO$_3$, $R$ = Sc, Y, Dy–Lu). The h-$R$MnO$_3$ family of materials are geometric improper ferroelectrics in that the spontaneous polarization $P$ along the hexagonal *c*-axis is a by-product of the trimerization of the MnO$_5$ polyhedra setting in at the structural phase transition around 1000 K (*30*). The primary order parameter, a 2D vector with length $Q$ and azimuthal angle $\phi$, describes the shift of apical oxygen atoms when the MnO$_5$ polyhedra tilt. The coupling between the polarization and trimerization leads to a firm clamping of ferroelectric DWs and structural anti-phase boundaries (*31, 32*). We show that the effective conductivity of DWs on the (001) surface at GHz frequencies is drastically higher than that at dc, while the effect is absent on surfaces with in-plane polarized domains. Theoretical calculations indicate that the observed behavior is consistent with the dielectric loss due to periodic sliding of the DW around its equilibrium position, i.e., the synchronized oscillation of local polarization and apical oxygen atoms. Our results represent a major milestone in understanding the structural DW dynamics in complex systems.

## II. Results

### A. Multi-mode microscopy on (001) YMnO$_3$

Our multi-mode imaging setup on a commercial atomic-force microscopy (AFM) platform is schematically illustrated in Fig. 1A. We first discuss the results on the (001) surface of as-grown YMnO$_3$ samples. With no corresponding topographic features, the cloverleaf-like domain patterns are vividly seen in the out-of-plane piezo-response force microscopy (PFM) data in Fig. 1B. Due to its semiconducting band gap of ~ 1.5 eV and slight p-doping from interstitial oxygen (*28*) during the growth, the as-grown YMnO$_3$ in our experiment shows a room-temperature conductivity $\sigma_{bulk}$ of ~ $0.3 \times 10^{-3}$ S/m (*31, 33, 34*) (Supplementary Materials Fig. S1). This bulk conduction is further modulated by the different surface band bending between up- and down-polarized domains, giving rise to the domain contrast in the C-AFM image under a tip bias of -5 V (*31, 34*). In contrast to the uncharged walls in other ferroelectrics (*16-18*), it was found that the charge-neutral DWs on the (001) YMnO$_3$ surface are more resistive than the adjacent domains (*31*). Since the paraelectric phase of YMnO$_3$ is more insulating than its ferroelectric phase, it was suggested that the dc behavior of the DWs closely resembles the corresponding high-temperature high-symmetry states (*31*).

Different from the dc C-AFM, the scanning impedance microscope (SIM) (*35, 36*) working at $f$ = 1 GHz measures the local complex permittivity of the material with a spatial resolution of ~ 100 nm determined by the tip diameter $d$ (Fig. S2). The input excitation power is on the order of 10 μW (*35*), corresponding to a low GHz tip voltage of ~ 0.1 V. And the electric field ~ 10 kV/cm at the tip apex is too small to cause ferroelectric switching of the YMnO$_3$ domains (*26, 27, 31*). The output SIM-Re and SIM-Im signals are proportional to the real and imaginary parts of the tip-sample admittance, respectively. The SIM images acquired on the same area as above are also displayed in Fig. 1B. Strikingly, while the DWs are the least conductive objects on the (001) surface at zero frequency, they exhibit much higher SIM signals than the bulk domains, as seen from the line profiles in Fig. 1C. Note that since the tip diameter is much larger than the size of the vortex cores (*37*), the high SIM signals at the cores may be a resolution-limited effect due to the summation of adjacent DW signals. We will not, therefore, analyze the SIM data on the vortices.

In order to interpret the SIM data as physical quantities, we use finite-element analysis (FEA) to simulate the DW response at GHz frequencies (*36*). Near the surface of plate-like YMnO$_3$ crystals, the DWs tend to be perpendicular to the surface for a depth ($h$) of several micrometers (*38*), i.e., $h \gg d$. As a result, we can model the DW as a vertical narrow slab sandwiched between adjacent domains. To compare results at different frequencies, we characterize the total dielectric loss at the DWs, including both contributions from mobile carriers and DW dynamics, by the effective DW ac conductivity $\sigma_{DW}^{ac}$. Detailed procedures of the FEA modeling are found in Fig. S3. The simulated SIM signals as a function of $\sigma_{DW}^{ac}$ are plotted in Fig. 1D. Compared with the FEA result, the measured DW contrast corresponds to $\sigma_{DW}^{ac} \approx 400$ S/m, which is 5 ~ 6 orders of magnitude higher than $\sigma_{DW}^{dc}$. The electrical response of YMnO$_3$ DWs is therefore qualitatively different from that of the ferroelectric lead zirconate (PZT) (*39*) and the magnetic insulator Nd$_2$Ir$_2$O$_7$ (*40*), where the DWs are more conductive than domains at both dc and GHz frequencies.

## B. Control experiments on other h-$R$MnO$_3$

In order to investigate the generality of the observed behavior, we have also performed control experiments on other h-$R$MnO$_3$ samples. As shown in Fig. 2A, the same DW contrast in SIM data is seen on the (001) surface of ErMnO$_3$, suggesting that the effect is insensitive to the variation of rare-earth elements. The situation, however, is rather different on h-$R$MnO$_3$ surfaces with in-plane polarized domains. In Fig. 2B, no DW contrast in the SIM images is observed on the cleaved (110) HoMnO$_3$ crystal. In addition, we cut two pieces from the same HoMnO$_3$ crystal and polished the (001) surface for one sample and the (100) surface for the other. As displayed in Fig. S4, the appearance and absence of SIM contrast of DWs are again observed on the (001) and (100) surfaces, respectively.

DWs inside h-$R$MnO$_3$ form a complex network that permeates the bulk of the material (*41*). On the (110) and (100) surfaces with in-plane polarization, charged DWs are stabilized by the interlocking of the ferroelectric and antiferrodistortive orders. Previous C-AFM work on these surfaces showed that $\sigma_{DW}^{dc}$ varies continuously as the neighbouring domains change from 'tail-to-tail' to 'head-to-head' configurations, presumably due to the accumulation or depletion of p-type carriers (*42*, *43*). On the other hand, only moderate conductance difference between domains and DWs (within an order of magnitude) is measured on these surfaces (*42*). Compared with the FEA results in Fig. 1D, the contribution due to mobile carriers on the charged walls ($\sigma_{DW}^{dc} \ll 1$ S/m) is still too small to be detected by the SIM. As a result, the missing DW contrast on the (110) and (100) surfaces at $f = 1$ GHz indicates that the contribution from dipolar loss to the local energy dissipation is also negligible in crystal planes parallel to the polarization axis.

## C. Frequency-dependent DW response

To further explore the unusual ac response of DWs on (001) YMnO$_3$, we construct multiple SIM electronics to cover a broad spectrum ranging from $10^6$ to $10^{10}$ Hz (Fig. S5). Selected SIM images with clear DW contrast are shown in Fig. 3A-D (more in Fig. S6). Due to the different settings such as input power, amplifier gains, and impedance-match sections of the electronics, the absolute SIM signals cannot be directly compared between different frequencies. Moreover, the SIM output is strongly dependent on the condition of the tip apex, as evident from the simulation results in Fig. 3E. We have therefore taken the ratios between SIM-Re and SIM-Im signals, which not only cancel out the circuit-dependent factors but also show much weaker dependence on the tip diameter (Fig. 3F), for quantitative analysis. As shown in Fig. S7, repeated line scans were also taken across several DWs to improve the signal-to-noise ratio at each $f$. In Fig. 3G, the SIM-Re/Im data are plotted together with the constant-$\sigma_{DW}^{ac}$ contours from the FEA simulation. Within the experimental errors in Fig. 3H, the effective DW conductivity rises rapidly from nearly zero at dc to ~ 500 S/m above 1 GHz and develops a feature not inconsistent with a broad peak around 3 ~ 5 GHz, although the cut-off frequency at 10 GHz prevents us from resolving the full resonance-like peak. Given the very small contribution from Drude conduction of mobile carriers, it is obvious that the bound-charge motion at the DWs, which is microscopically equivalent to the vibration of DW position, on the (001) surface of h-$R$MnO$_3$ is responsible for the pronounced ac loss observed in our experiment.

### D. Theoretical analysis of DW Dynamics in h-$R$MnO$_3$

The starting point to analyze the lattice dynamics in h-$R$MnO$_3$ is to understand its non-uniform trimerization textures, which were first explained within the long-wavelength Landau theory, using parameters extracted from *ab initio* calculations (*44*). Density functional theory (DFT) calculations predict the lowest optical phonon at ~ 2 THz (*44*, *45*), well above the characteristic frequency in our experiment. On the other hand, the presence of DWs breaks the continuous translational symmetry and introduces a mode associated with the periodic DW sliding around its equilibrium position, whose energy approaches zero in the continuum model. When the DW width $\lambda$ is comparable to the lattice constant $a$, the sliding mode acquires a gap due to the discrete translational symmetry of the lattice. The essential physics of DW dynamics can be captured by a simplified one-dimensional model Hamiltonian:

$$H = \sum_r \frac{\left(\dot{A}_r\right)^2}{2m} + b\left(A_r^2 - 1\right)^2 + \frac{c}{2}\left(A_r - A_{r+\delta}\right)^2 - EA, \qquad [1]$$

where the four terms represent the kinetic energy of the local mode $A_r$ at site $r$ with mass $m$, the local double-well potential, the nearest-neighbor interaction, and the interaction between the local mode and oscillating external field $E$, respectively. In the equilibrium, the center of the DW locates in between two adjacent sites, as shown in Fig. 4A, so that the mode amplitude at every site is close to the minimum of the ferroelectric double-well potential. When moving to a neighboring unit cell, the DW passes through an intermediate configuration in Fig. 4B, where it is centered at the site. The mode amplitude of this configuration corresponds to a maximum of the potential and a high total energy. The energy difference between these two configurations gives rise to the Pierls-Nabarro barrier (*46*, *47*) in Fig. 4C that needs to be overcome in order to flip the polarization of a unit cell and move the DW to the adjacent cell. In our SIM experiment, the excitation tip voltage, thus the external field $E$, is too low to cause any ferroelectric switching. As a result, we only need to consider the linear response of this model Hamiltonian.

The solution of model [1] reveals two non-dispersive modes localized perpendicular to the DW plane, along with the continuous spectrum of bulk phonons, as shown in Fig. 4D. The lowest-energy mode (Fig. 4E) corresponds to oscillations of the DW position, whereas the higher-energy breathing mode (Fig. 4F) corresponds to oscillations of the DW width. As seen in Fig. 4G, the frequency of the DW sliding mode rapidly decreases with increasing $\lambda$ and approaches zero for $\lambda \gg a$, consistent with the free DW sliding in the continuum theory. The nearly-exponential dependence of the resonance frequency on $\lambda$ makes it difficult to calculate the frequency precisely from an approximate model, but it is still possible to estimate it by the order of magnitude. This simple model thus illustrates how the low-energy GHz-scale mode emerges from the THz phonon spectrum.

Now we extend the above model to incorporate the principal trimerization amplitude ($Q$), angle ($\phi$), and polarization ($P$) modes in $YMnO_3$. Since the DW vibration occurs at a frequency much lower than the optical phonons, it is within the error bars of the conventional frozen-phonon calculations (*48*). To provide a quantitative estimate, we use the discretized version of the model in Ref. (*44*), where the effective masses and inter-site couplings are extracted from the phonon dispersion calculated within DFT. Detailed calculations on the full model Hamiltonian of a 120-site supercell with a DW centered in the middle are included in Fig. S8. The DW energy centered at the Mn sites is lower than that in between the sites, again giving rise to the washboard-like Pierls-Nabarro barrier. The atomistic view of $YMnO_3$ in Fig. 5A shows the synchronized oscillation of apical oxygen atoms and the local polarization during the periodic DW sliding. In Fig. 5B, the ground-state configuration by minimizing the model Hamiltonian is plotted for $Q$, $\phi$, and $P$. The calculated phonon dispersion can be visualized by the phonon spectral function in Fig. 5C, where the damping is estimated from optical experiments (*49*).

Similar to the simplified model, the spectrum contains the usual dispersive phonons originated from the $\Gamma_2^-$ and $K_3$ modes in the P6$_3$/mmc space group (*45*) and non-dispersive branches, the lowest of which corresponds to the localized DW oscillating mode. In Fig. 5D, the order parameter oscillation amplitudes at the lattice sites are plotted for both a regular phonon and the localized mode. Note that our 1D model calculations only capture the bottom of the phonon bands, i.e. phonons with zero wave vector in the DW plane. In real 3D crystals, the vibration of local polarization and oxygen atoms can still propagate with non-zero wave vectors within the DW plane, resembling the acoustic waves in elastic media. The bandwidth of such a DW-acoustic-wave mode is determined by the inter-site coupling strength, and is therefore of the same order as the bandwidth of the $\Gamma_2^-$ and $K_3$ modes in the bulk.

## III. Discussion

Comparisons between our experimental and theoretical results strongly suggest that the ac response of the h-$R$MnO$_3$ DWs is associated with the DW oscillation mode. First, for DWs on the (001) surface, the vertical component of the oscillating *E*-field from the SIM tip is aligned with either up- or down-polarized domains in each half-cycle, leading to the periodic motion of the apical oxygen atoms coupled with local polarization and thus the ac dielectric loss. In contrast, the coupling between the tip *E*-fields and the DW motion on the (110) and (100) surfaces is negligible due to *E* being mostly orthogonal to *P*. Such a 'selection rule' is indeed seen in Fig. 2 and Fig. S4. Secondly, while the error bars of a full *ab initio* calculation for all 3×30 h-$R$MnO$_3$ phonons can mask the low-energy mode, our model Hamiltonian taking the essential $Q$, $\phi$, and $P$ modes into account provides an estimate of its frequency that indeed falls into the GHz regime. Further development of the model, such as the inclusion of the next-nearest-neighbour interaction and more phonon modes, may allow a more quantitative comparison with the experiment. Thirdly, our generic analysis of the low-energy DW vibration may explain the dielectric dispersion observed in many ferroelectrics (*7-12*). Future experiments on materials with a characteristic frequency well within the range of our SIM, e.g., 0.1 ~ 1 GHz, will allow us to fully resolve the resonance-like peak (similar to Fig. 3H), which contains important information such as the DW effective mass. Finally, our model calculations suggest the possible presence of an acoustic mode localized perpendicular to the wall but free to propagate within the DW plane. If confirmed by future experiments, this low-energy excitation, in analogy to the magnonic wave traveling along the magnetic DWs (*50*) and surface acoustic wave traveling on the surface of piezoelectric materials (*51*), may be exploited for nanoelectronic applications.

To summarize, by using broadband impedance microscopy, we have observed the drastic increase of effective DW conductivity from dc to microwave frequency on the (001) surfaces of hexagonal manganites, while the effect is absent on surfaces with in-plane polarized domains. First-principles and model calculations indicate that the DW oscillation, rather than the presence of free carriers, is responsible for the ac energy loss and selection rules. Ferroelectric DWs, with their own rich excitations, thus offer a new playground to explore emergent interfacial phenomena that are not present in bulk domains.

## Materials and Methods

**Material preparation.** Plate-like single crystals of hexagonal h-$R$MnO$_3$ with a few millimeters in-plane size were grown by a flux method with a mixture of h-$R$MnO$_3$ polycrystalline powder and Bi$_2$O$_3$ (the molar ratio of 1:6), slowly cooled from 1280 °C to 970 °C at the rate of 2 °C/h. In order to prepare type-I ferroelectric domains of h-$R$MnO$_3$, single crystals of h-$R$MnO$_3$ were annealed at 1180 °C, above its $T_C$, in Ar atmosphere and slowly cooled to 1000 °C in 2 hours, and then quenched to room temperature to avoid surface oxidation at lower temperatures.

**Scanning Impedance Microscopy.** The multi-mode imaging experiments, including SIM, PFM, and C-AFM, were performed in an AFM platform (XE-70) from Park Systems. Customized electronics were used for impedance imaging at frequencies above 50 MHz. The HF2LI lock-in amplifier from Zurich Instruments was configured to perform SIM at frequencies below 50 MHz. The DLPCA-200 current amplifier from FEMTO Inc. was used for the C-AFM imaging. The micro-fabricated shielded probes are commercially available from PrimeNano Inc. Finite-element modeling was performed by using the commercial software COMSOL 4.4.

## Supplementary Materials

Supplementary Materials for this article are available online.

## Acknowledgements

We thank Weida Wu for helpful discussions. The SIM work was supported by the U. S. National Science Foundation, Division of Materials Research, Award #1649490. The instrumentation was supported by the U. S. Army Research Laboratory and the U. S. Army Research Office under grant number #W911NF1410483. The work at Rutgers is funded by the Gordon and Betty Moore Foundation's EPiQS Initiative through Grant GBMF4413 to the Rutgers Center for Emergent Materials.

## Author contributions

S.-W.C. and K.L. conceived and designed the experiments. R.H. grew the materials. X.W., L.Z. and Y.R. constructed the electronics and performed the SIM experiment and numerical analysis. U.P. and S.A. performed the theoretical studies. X.W., S.A., and K.L. wrote the initial draft of the paper. All authors were involved in the discussion of results and edited the manuscript.

## Competing interests

The authors declare no competing financial interests.

## Data and materials availability

All data needed to evaluate the conclusions in the paper are present in the paper and/or the Supplementary Materials. Additional data related to this paper may be requested from the authors.

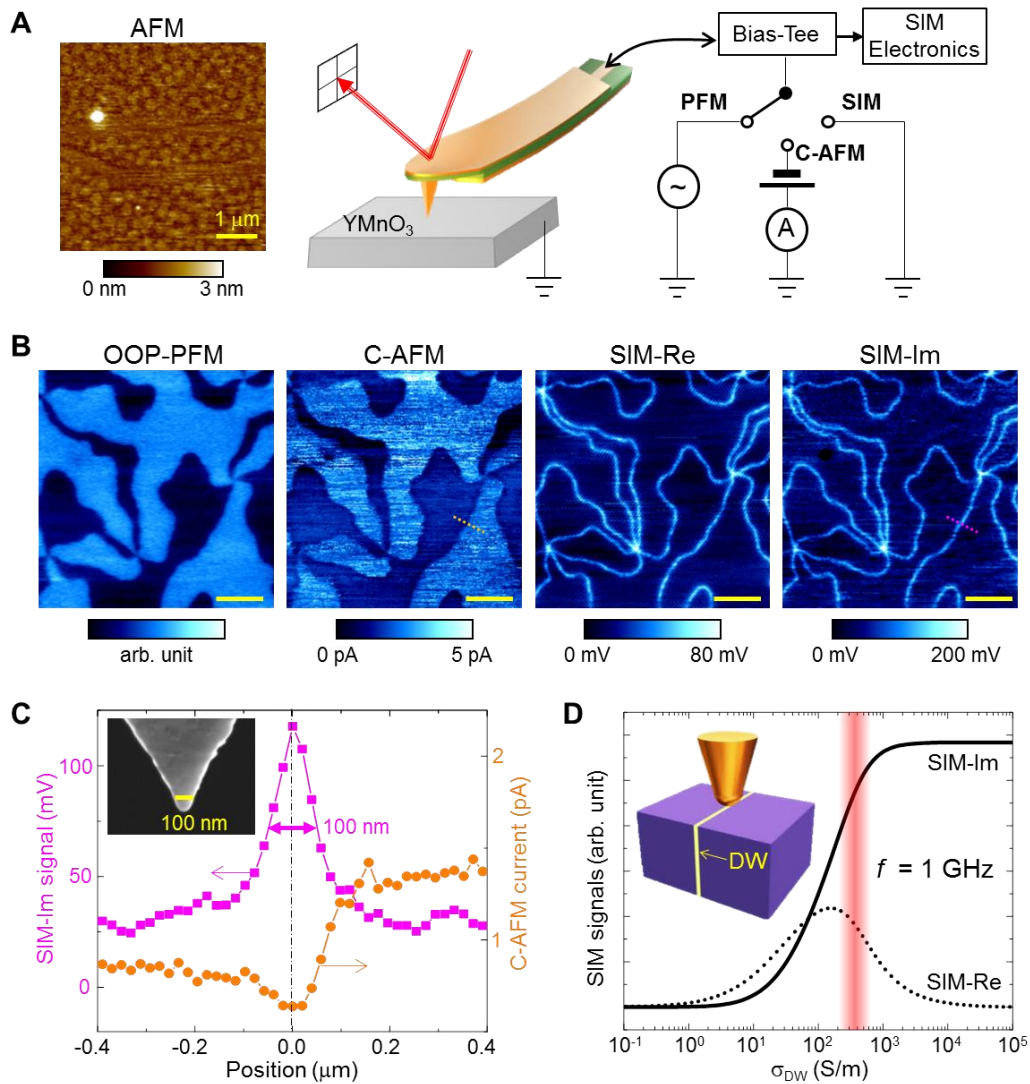

**Figure 1. Multi-mode microscopy on (001) YMnO$_3$.** (**A**) Schematic of the experimental setup. The shielded cantilever probe is connected to the SIM electronics via a bias-tee, through which a low-frequency ac voltage (95 kHz, 5 V) for PFM or a dc bias (−5 V) for C-AFM can be applied to the tip. The AFM image on the left shows the surface topography of (001) YMnO$_3$. (**B**) Out-of-plane (OOP) PFM, C-AFM, SIM-Re, and SIM-Im ($f$ = 1 GHz) images acquired on the same area. All scale bars are 1 μm. (**C**) SIM-Im (purple) and C-AFM (orange) line profiles across a single domain wall centered at position 0.0 μm and labeled as dashed lines in **B**. The full-width-half-maximum of 100 nm is comparable to the tip diameter, as shown in the scanning electron microscopy (SEM) image in the inset. (**D**) Simulated SIM signals as a function of the effective DW conductivity. The measured DW signals with a ratio of SIM-Re/Im ~ 0.4 (shaded in red) are consistent with $\sigma_{DW}$ ~ 400 S/m. The inset shows the tip-sample geometry for the FEA.

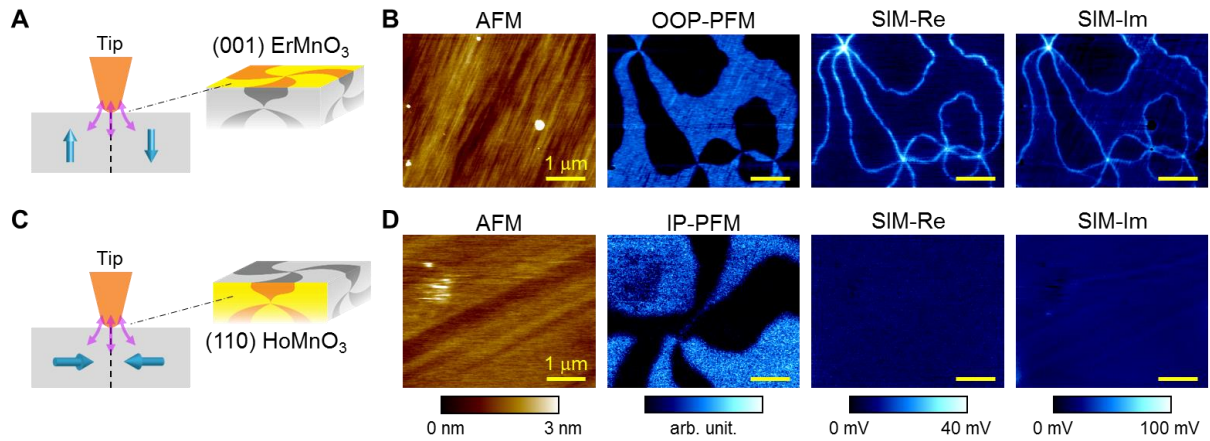

**Figure 2. SIM experiments on other h-$R$MnO$_3$.** (**A**) Schematic representation of the tip electric fields (purple) and the out-of-plane polarization (blue) on the highlighted (001) ErMnO$_3$ surface. (**B**) AFM, out-of-plane PFM, SIM-Re, and SIM-Im ($f$ = 1 GHz) images acquired on (001) ErMnO$_3$. Clear DW contrast can be seen in the SIM data. (**C**) and (**D**) are the same as (**A**) and (**B**) except that the schematic and the data are for (110) HoMnO$_3$, showing clear domain contrast in the in-plane PFM but no DW contrast in the SIM images. All scale bars are 1 μm.

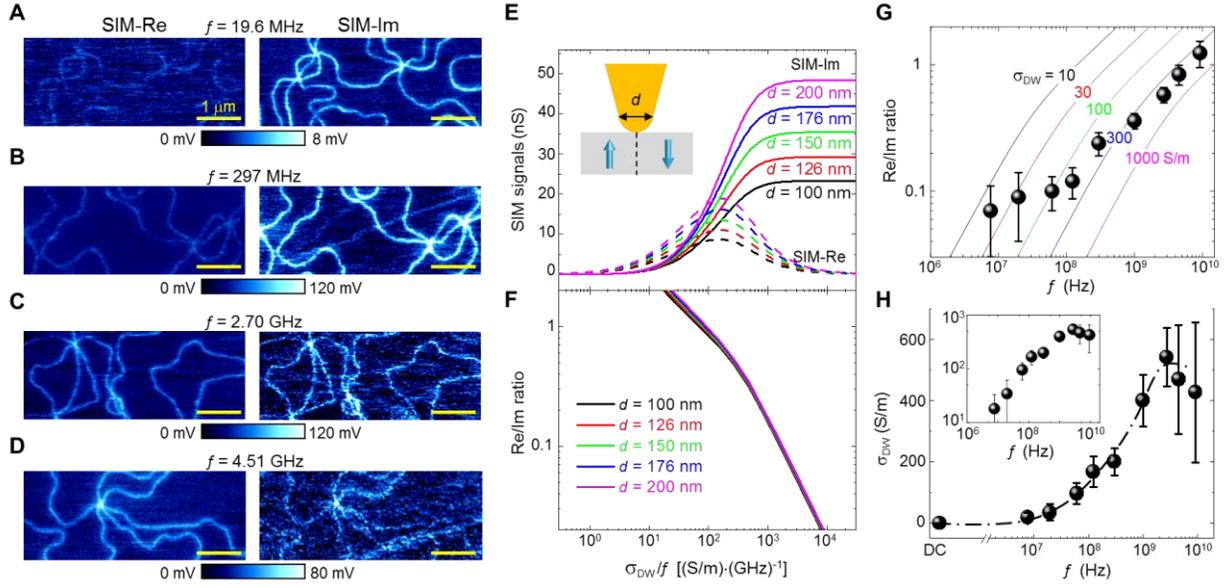

**Figure 3. Frequency dependent DW response.** (**A-D**) SIM images on (001) YMnO$_3$ at selected frequencies. All scale bars are 1 μm. (**E**) Simulated SIM signals and (**F**) SIM-Re/Im ratios for different tip diameters, showing the weak dependence on the exact tip condition when the Re/Im ratio is calculated. Note that the *x*-axis is $\sigma_{DW}/f$, i.e., the simulation is invariant when $\sigma_{DW}$ is scaled by the frequency. (**G**) SIM-Re/Im ratio of the DW signals as a function of *f* in a log-log plot. The constant $\sigma_{DW}$ contours at 10, 30, 100, 300, 1000 S/m are also plotted in the graph. (**H**) *f*-dependent $\sigma_{DW}$ of the (001) YMnO$_3$ DWs. The dash-dot line is a guide to the eyes. The inset shows the same data in the log-log scale.

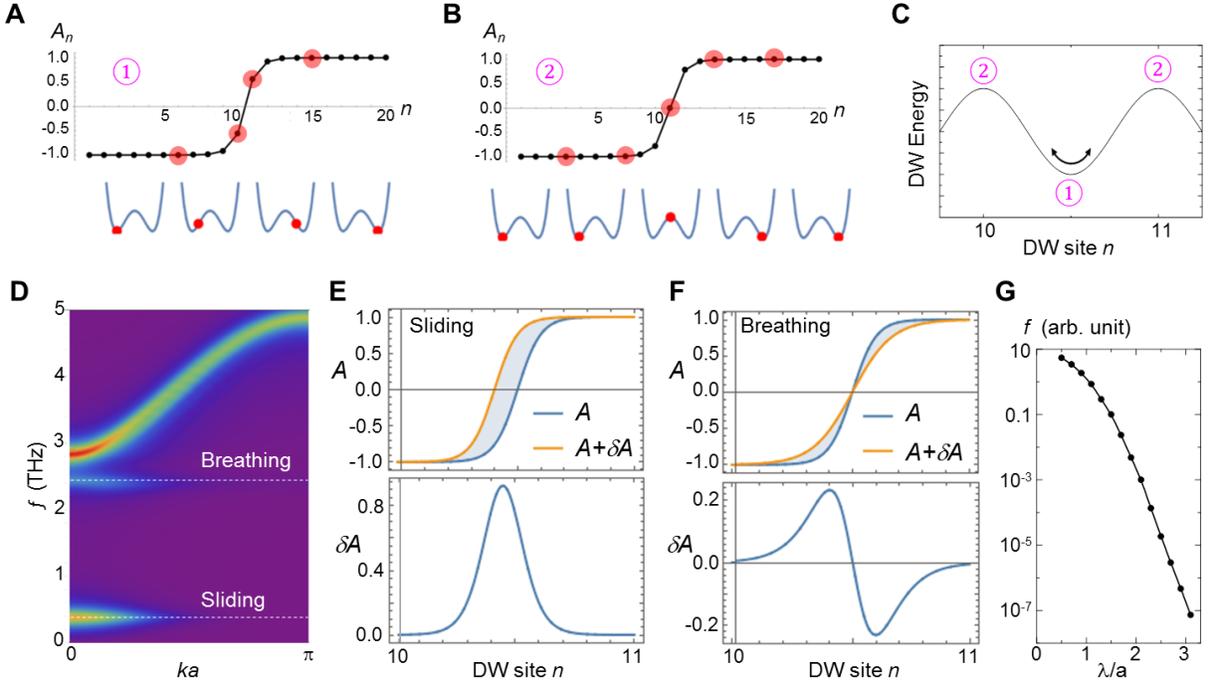

**Figure 4. Periodic DW sliding in the simplified model.** (**A**) Ground-state configuration of the simplified Hamiltonian (1), with the DW centered between two Mn sites. (**B**) A high-energy configuration when the domain wall is centered at a Mn site. The schematics in (**A**) and (**B**) show the corresponding on-site energies in the double-well potential. (**C**) Washboard-like potential when the center of the DW slides across different sites. (**D**) Phonon spectral function in this simple model, showing the non-dispersive sliding mode at the lowest energy, the breathing mode at a higher energy, and the dispersive bulk phonon branch. (**E**) Mode texture for a lateral shift of the DW position (top) and the corresponding DW sliding mode (bottom). (**F**) Mode texture for an increase of the DW width (top) and the corresponding DW breathing mode (bottom). (**G**) Dependence of the DW oscillation frequency on its width.

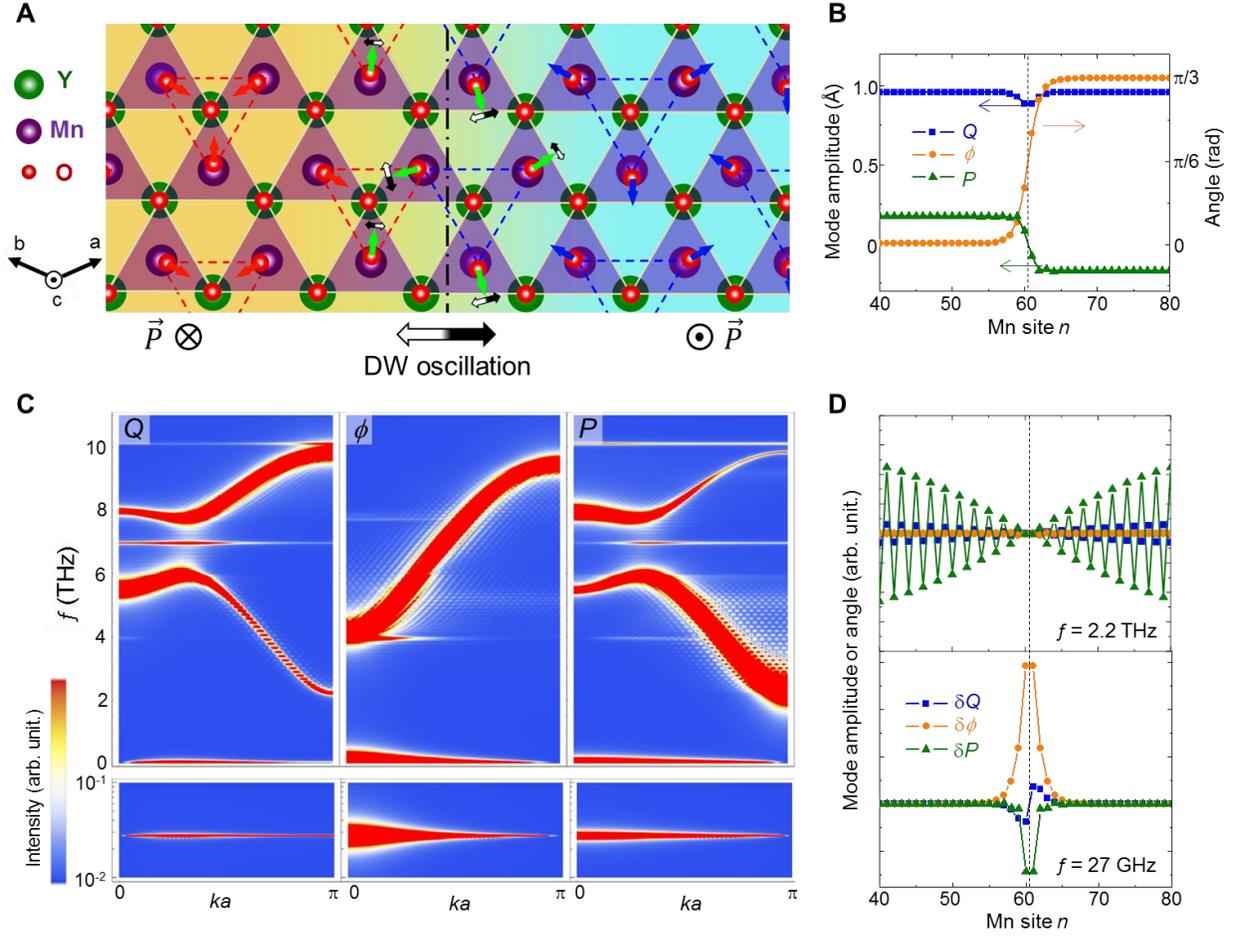

**Figure 5. DW dynamics revealed by first-principles-based model calculations.** (**A**) Atomistic view of YMnO$_3$ in the (001) plane across the interlocked antiphase boundary and ferroelectric DW. The MnO$_5$ polyhedra are shaded in purple. The displacements of apical oxygen atoms in the down-domain (left), DW (middle), and up-domain (right) regions are displayed by red, green, and blue arrows, respectively. The trimers are indicated by dashed triangles. The black and white double-headed arrows illustrate the amplitudes and directions of the periodic DW sliding. (**B**) Ground-state configuration of the three order parameters across the DW obtained by minimizing the model Hamiltonian. (**C**) Phonon spectral function projected to the $Q$, $\phi$, $P$ modes. The ripples are due to the finite size (120 sites) of the super-cell. The lower panels in the log scale show the spectral intensity of the low-energy non-dispersive branch. (**D**) Real-space oscillation of $\delta Q$, $\delta\phi$, and $\delta P$ for (top) a regular dispersive phonon at the THz-range and (bottom) the localized DW sliding mode at the GHz-range.

# Supplementary Materials

## Low-energy Structural Dynamics of Ferroelectric Domain Walls in Hexagonal Rare-earth Manganites

Xiaoyu Wu[1], Urko Petralanda[2], Lu Zheng[1], Yuan Ren[1], Rongwei Hu[3], Sang-Wook Cheong[3*], Sergey Artyukhin[2*], Keji Lai[1*]

1. Department of Physics, University of Texas at Austin, Austin, Texas 78712, USA

2. Quantum Materials Theory, Istituto Italiano di Tecnologia, Genova, Italy

3. Rutgers Center for Emergent Materials and Department of Physics and Astronomy, Rutgers University, Piscataway, NJ 08854

* Correspondence should be addressed to S.-W. C. (sangc@physics.rutgers.edu), S. A. (Sergey.Artyukhin@iit.it) and K. L. (kejilai@physics.utexas.edu)

**Section 1: DC conductivity of h-*R*MnO$_3$.**

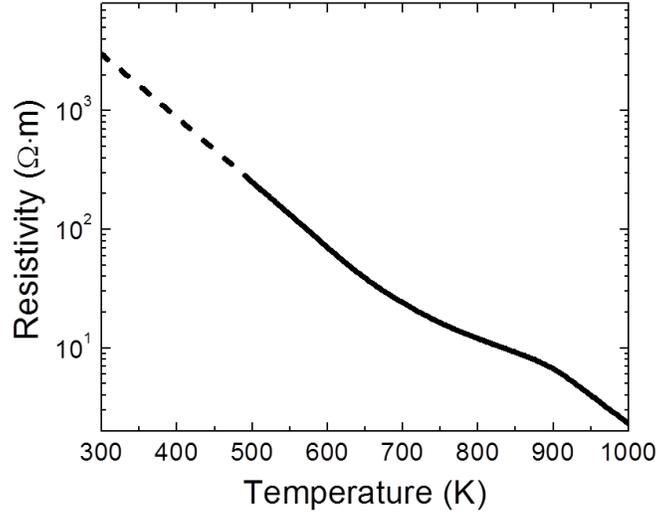

**Figure S1. Measurement of the dc resistivity of YMnO$_3$.**

The dc resistivity of our single-crystal YMnO$_3$ samples is measured by a four-probe method using a Keithley source meter. The electrical contacts are made by curing the gold paste at high temperatures between 600 K and 800 K. The results from 1000 K to 500 K are shown in a solid curve because good Ohmic contacts are achieved in this temperature range. The kink at ~ 850 K is likely due to the formation of oxygen interstitials, which effectively dope the surface of the crystals (*28*). For temperatures below 500 K, the contact resistance becomes substantially large and the I-V curves are no longer Ohmic-like. An extrapolation from ~ 600 K to ~ 300 K indicates a room-temperature dc resistivity of ~ 3×10$^3$ Ω·m, or the conductivity of ~ 0.3×10$^{-3}$ S/m, which is consistent with the number (on the order of 10$^{-3}$ S/m) quoted in the literature (*31, 33, 34*). Future experiments using samples with improved Ohmic contacts and instruments with higher input impedance may provide a more quantitative room-temperature dc resistivity of the YMnO$_3$ crystals.

**Section 2: SIM electronics and the calibration process.**

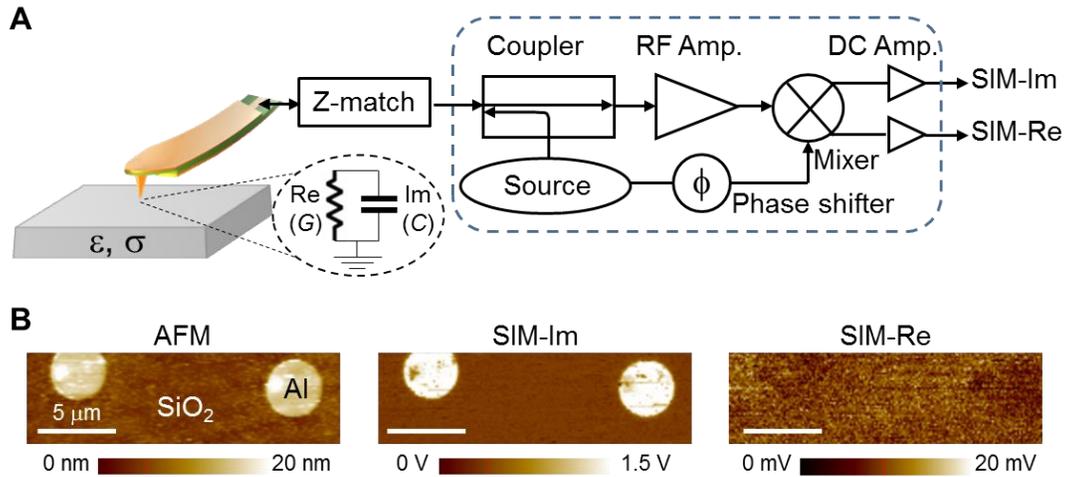

**Figure S2. SIM electronics and the calibration process.** (**A**) Schematics of the SIM electronics. (**B**) AFM and SIM images ($f$ = 1 GHz) on the standard sample with Al dots on the $SiO_2$/Si substrate. All scale bars are 5 μm.

The schematic diagram of the SIM electronics is shown in Fig. S2A. The radio-frequency (rf) signal is sent to the tip through an impedance-match section, which is a quarter-wave resonator with a small reflection coefficient ($S_{11}$ ~ -20 dB) at $f$ = 1 GHz. All our measurements are performed in the linear regime, i.e., the SIM output is proportional to the tip voltage. With a typical excitation power of 10 μW, a simple transmission-line analysis taking into account the cable loss show that the rf voltage at the tip is on the order of 0.1 V. The reflected wave is amplified and demodulated to form the two output channels. The SIM-Re and SIM-Im signals are proportional to the real (in-phase) and imaginary (out-of-phase) parts of the tip-sample admittance, respectively.

An important step before the actual measurement is the channel alignment using standard samples such as Al dots on the $SiO_2$/n++Si substrate. On this sample, the tip-sample capacitance is different between the ~4 nm native $Al_2O_3$ (on top of the metallic Al) and the 100 nm $SiO_2$ (on top of the metallic n++ Si). As a result, the contrast should appear only in the SIM-Im channel. This condition is met by adjusting the phase shifter in front of the mixer to nullify the contrast in the SIM-Re channel. As shown in Fig. S2B, we can obtain good channel alignment at 1 GHz, with a residual SIM-Re/Im ratio < 1%. Even towards the frequency limits of our broadband SIM (1 MHz and 10 GHz), an uncertainty within 5% is feasible. Since the SIM-Re/Im data points in Fig. 3 are all larger than 0.05, the results are genuine effects rather than instrumental artifacts.

**Section 3: Finite-element analysis of the tip-sample interaction.**

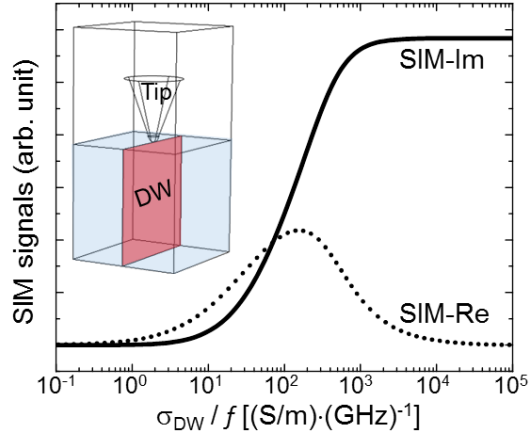

**Figure S3. Finite-element analysis of the tip-sample interaction.** Finite-element analysis (FEA) of the SIM signals as a function of $\sigma_{DW}$, which is scaled by the frequency in unit of GHz. The 3D modeling geometry is shown in the inset.

The FEA results are shown in Fig. S3. Here SIM-Re and SIM-Im signals are directly proportional to the real and imaginary parts of the tip-sample admittance (inverse impedance), which can be computed by the software COMSOL 4.4 (*36*). Note that the SIM signals saturate at both the low (below 1 S/m at 1 GHz) and high (above $10^4$ S/m at 1 GHz) conductivity limits. Due to the small bulk conductivity $\sigma_{bulk} \sim 10^{-3}$ S/m, the SIM response on the bulk domains is in the insulating limit. The quasi-static simulation is invariant when the effective DW conductivity $\sigma_{DW}$ is scaled by the frequency, i.e., the curves shift to higher $\sigma_{DW}$ at higher frequency and vice versa. A tip diameter (*d*) of 100 nm was used in the modeling, consistent with the SEM image at the tip apex and the line profiles in Fig. 1C. The domain wall width ($\lambda$) on the order of 1 nm was reported in previous transmission electron microscopy studies (*37, 52*). We assign $\lambda = 2$ nm so that it is easy to generate a dense mesh in the simulation. In fact, since $d \gg \lambda$, the modeling result is invariant with respect to the effective sheet conductance $S_{DW} = \sigma_{DW} \cdot \lambda$. For instance, $\sigma_{DW} = 500$ S/m and $\lambda = 2$ nm lead to $S_{DW} = 1$ μS·sq and an effective sheet resistance $R_{DW} = 1/S_{DW} = 1$ MΩ/sq. Due to the relatively small static dielectric constant $\varepsilon_s = 17$ of the improper ferroelectric h-$R$MnO$_3$ (*53*), it is unlikely that a possible dipolar relaxation process (*54*) ($\varepsilon'$ drops from $\varepsilon_s$ to $\varepsilon_\infty < \varepsilon_s$ and $\varepsilon''$ peaks at $(\varepsilon_s + \varepsilon_\infty)/2$) within our frequency span will substantially affect the simulation result. As a result, we have chosen to interpret the observed impedance contrast in terms of a single parameter, i.e., the effective ac conductivity of the DWs, while making the reasonable assumption of *f*-independent dielectric constant $\varepsilon = 17$ for both domains and DWs.

## Section 4: SIM data on polished HoMnO$_3$ samples.

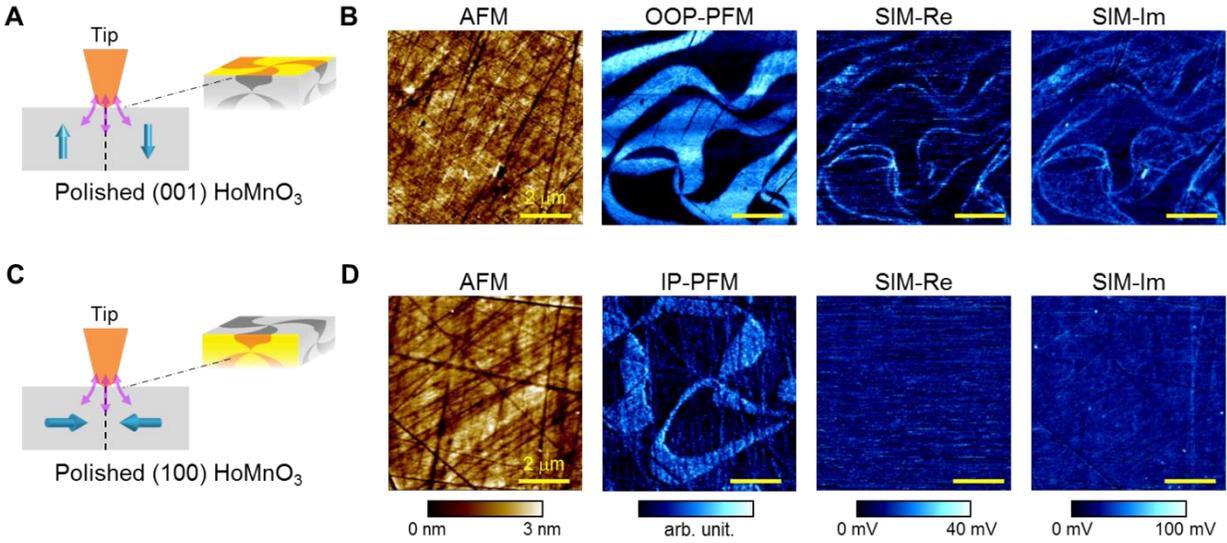

**Figure S4. SIM data on polished HoMnO$_3$ samples.** (**A**) Schematic and (**B**) AFM, out-of-plane PFM, and SIM images on the polished (001) surface of HoMnO$_3$ single crystal. (**C**) Schematic and (**D**) AFM, in-plane PFM, and SIM images on polished (100) surface cut from the same sample. All scale bars are 1 μm.

In addition to the as-grown YMnO$_3$, ErMnO$_3$ and cleaved HoMnO$_3$ samples presented in the main text, we have also measured other samples to confirm that the results are common to the h-$R$MnO$_3$ family. In particular, we cut and polished two samples, one with (001) surface and the other with (100) surface, from one single piece of HoMnO$_3$. Note that the SIM data are usually of low quality for polished samples with inevitable scratches and possible damages on the surface. Nevertheless, the images in Fig. S4B and S4D clearly show the appearance and absence of DW contrast on the (001) and (100) surfaces, respectively. In addition, the SIM-Re/Im ratio of the (001) sample (0.40±0.10) is consistent with that on the YMnO$_3$ and ErMnO$_3$ samples. We therefore conclude that the observed DW signals do not depend on the choice of the rare-earth element $R$, and the DW contrast is most prominent on the $ab$-plane with out-of-plane polarizations (*31*) and absent on surfaces parallel to the $c$-axis with in-plane polarizations (*42, 43*).

**Section 5: SIM circuits and impedance match at different frequencies.**

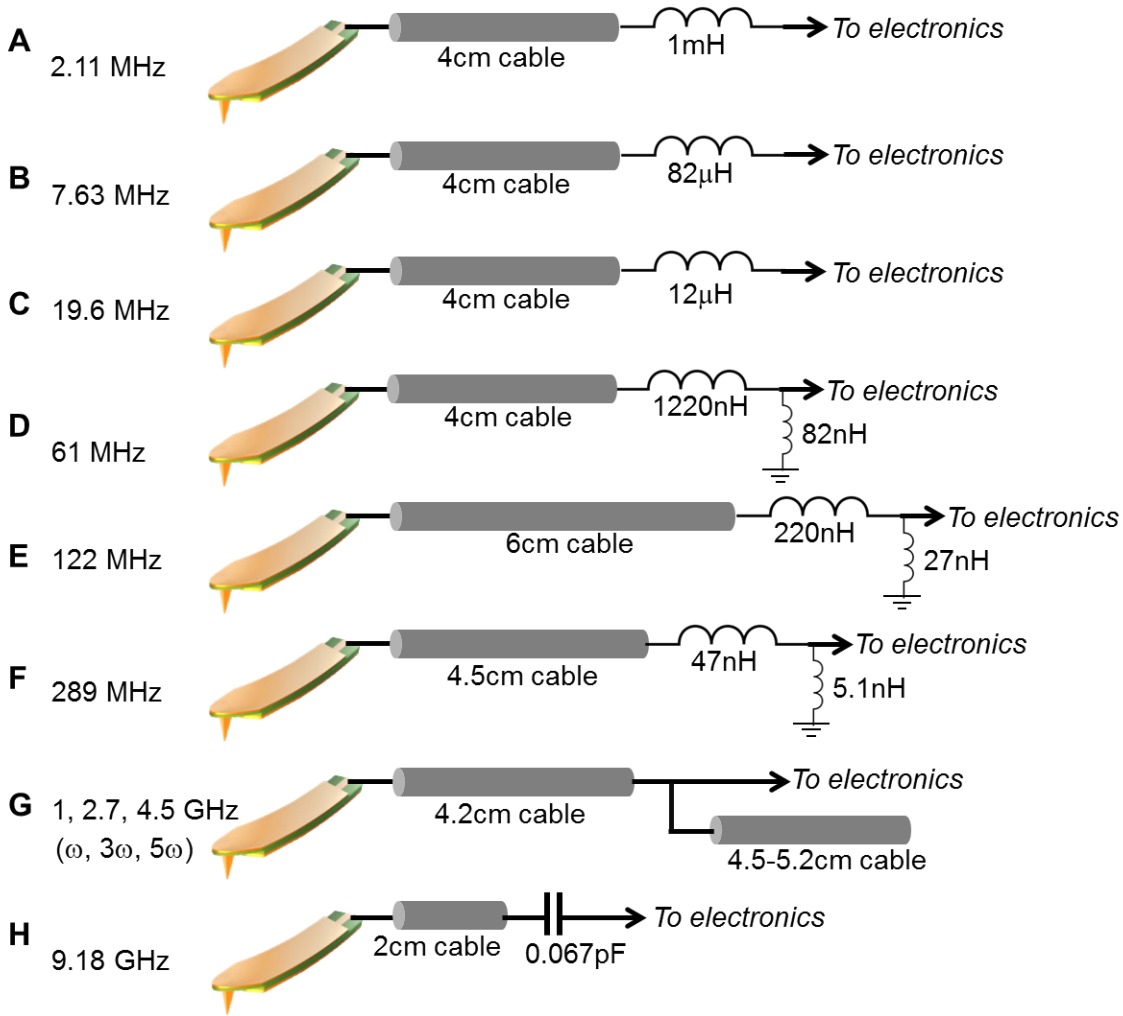

**Figure S5. Impedance-match sections at different frequencies.** (**A-H**) Impedance-match sections for all frequency points in this experiment.

We have constructed three sets of SIM electronics to obtain the best sensitivity around 100 MHz, 1 GHz, and 10 GHz, respectively. For frequencies below 50 MHz, we configured the HF2LI lock-in amplifier from Zurich Instruments to perform the SIM imaging. In addition, an impedance-match section (*36*) is needed at each frequency to route the tip to the 50 Ω transmission lines, as shown in Fig. S5. The impedance match is important to enhance the measurement sensitivity. It also adds another complication when comparing the absolute signal strengths at different frequencies. Again, the aforementioned method of taking the SIM-Re/Im ratio effectively cancels out this circuit-dependent factor and is therefore preferred for quantitative analysis of the DW signals.

**Section 6: More SIM data at various frequencies.**

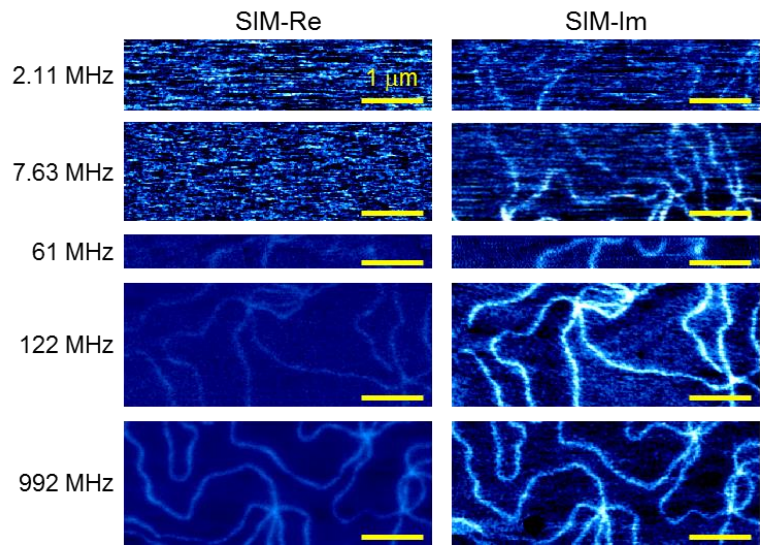

**Figure S6. SIM images at various frequencies.** All scale bars are 1 μm.

SIM images not shown in the main text (Fig. 3A) are included in Fig. S6. For each $f$, the SIM-Re and SIM-Im images are displayed with the same false-color scale. The absolute signals, however, cannot be compared between different electronics and therefore are not shown here.

**Section 7: Repeated line scans for improving the signal-to-noise ratio.**

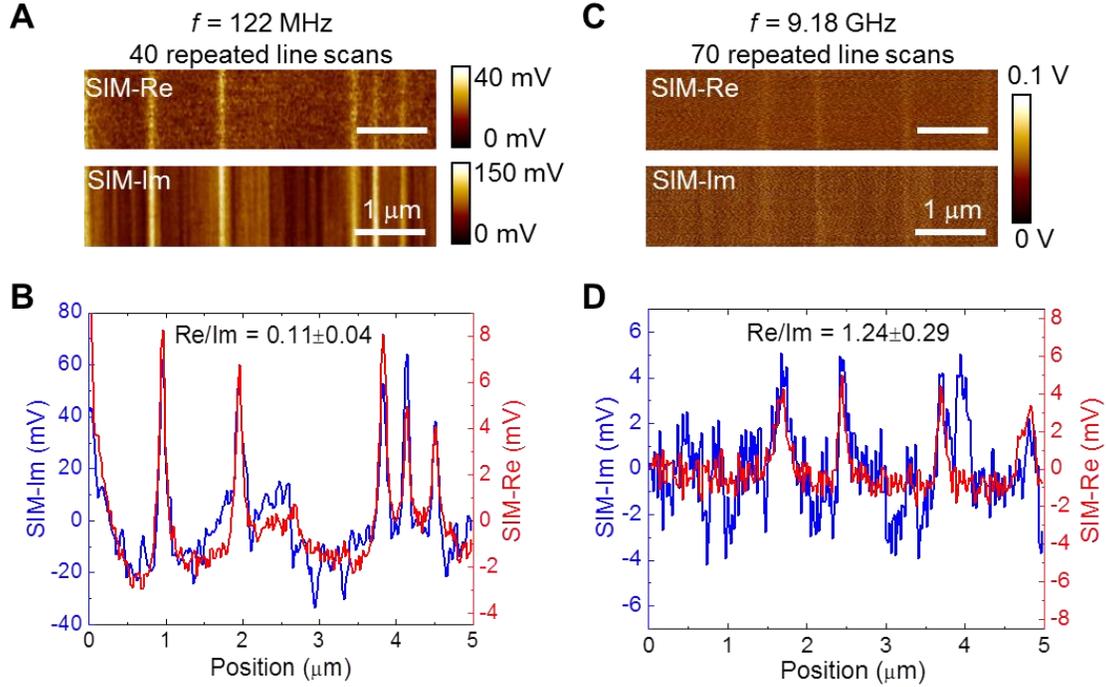

**Figure S7. SIM experiments with repeated line scans.** (**A**) Repeated SIM line scans at $f = 122$ MHz and (**B**) the corresponding averaged signals for calculating SIM-Re/Im with a better signal-to-noise ratio. (**C**) and (**D**) show the same results acquired at $f = 9.18$ GHz. All scale bars are 1 µm.

For the same capacitance contrast $\Delta C$, the admittance change ($2\pi f \cdot \Delta C$) becomes smaller at lower frequencies, which is thus harder to detect by the SIM electronics. The lowest frequency in this work is ~ 2 MHz, below which the signal-to-noise ratio (SNR) is too low for the imaging experiment. At high frequencies, the loss in the coaxial cables and the shielded cantilever sets our upper limit to be ~ 10 GHz, beyond which the SNR is again too low for imaging.

To extract the SIM-Re/Im ratio of the DW contrast, we perform repeated line scans to improve the SNR of the data. Figs. S7A and S7B show the results ($f = 122$ MHz) of 40 scans on the same line and the plot of averaged SIM signals, respectively. For the $f = 9.18$ GHz data shown in Figs. S7C and S7D, the DWs are hardly seen in the raw data due to the poor sensitivity. The line averaging method, however, provides adequate SNR such that the SIM-Re/Im ratio can be readily calculated.

**Section 8: Details of the full model calculations.**

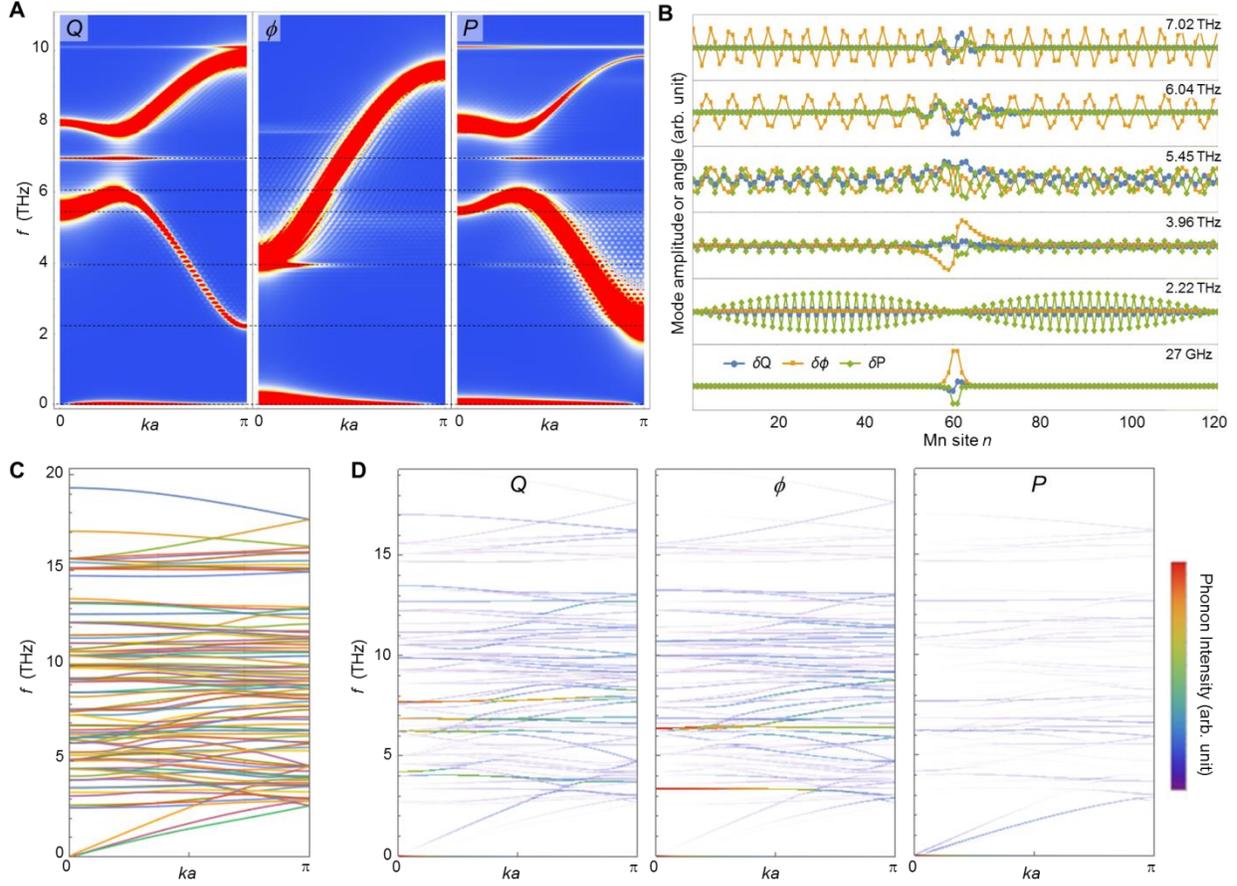

**Figure S8. First-principles-based model calculations.** (**A**) Phonon spectral function displayed in Fig. 5C. (**B**) Oscillation amplitudes of the local modes $Q$, $\phi$, and $P$ are shown for several characteristic phonons at frequencies labeled in **A** with the dashed lines. (**C**) Phonon dispersion of YMnO$_3$ in the low temperature P6$_3$cm structure. (**D**) From left to right: Contributions of the $Q$, $\phi$, and $P$ modes to the phonon dispersion. The lines are color coded with the relative intensity.

In improper ferroelectric YMnO$_3$ the trimerization and polarization are strongly coupled, therefore the structural (trimerization) antiphase boundaries are also ferroelectric DWs. In order to get a quantitative estimate for the sliding mode frequency, we extend the model to incorporate the principal modes, $Q$, $\phi$ and $P$, appearing at the trimerization transition. The structural changes across the domain walls and other non-uniform trimerization textures are primarily described by the modulation of these modes (*44*). We therefore use the discretized version of the model (*44*), involving all the relevant degrees of freedom, $Q$, $\phi$, and $P$, and replace the gradient terms with the corresponding interactions between the neighboring sites:

$$H = \sum_{n=1}^{L} \frac{\left(\dot{Q}_n\right)^2 + Q_n^2 \dot{\phi}_n^2}{2m_Q} + \frac{\left(\dot{P}_n\right)^2}{2m_P} - A_1\left(Q_n^2 - Q_0^2\right)^2$$
$$-c_1\left((Q_n - Q_{n+1})^2 + \frac{(Q_n + Q_{n+1})^2(\phi_n - \phi_{n+1})^2}{4}\right) \quad [\text{S1}]$$
$$-A_2 P_n^2 - c_2\left(P_n - P_{n+1}\right)^2 - g_1 Q_n^3 P_n \cos 3\phi_n - g_2 Q_n^2 P_n^2.$$

In order to approximate the realistic phonon dispersion, we use the corresponding interaction constants and mode masses to fit the phonon band dispersion and band center positions to those determined by the DFT calculations. To calculate the phonons within the model, the Hamiltonian is expanded around the equilibrium DW configuration, up to the terms, quadratic in deviations from the ground state DW, and the corresponding Euler-Lagrange equations are solved as an eigenvalue problem. The finite chain is used with the fixed boundary conditions. The resulting phonon modes with energies $E_n$ and amplitudes $x_{nm}^{(\alpha)}$ of the mode $\alpha$ ($\alpha$ = 1, 2, 3 for modes $Q$, $\phi$, and $P$) on the site $m$ can be visualized by plotting the phonon spectral function in Fig. S8A,

$$A_{\alpha\beta}(k,\omega) = \sum_n \delta(\omega - E_n)\langle k | x_{nm}^{(\alpha)} \rangle \langle x_{nm}^{(\beta)} | k \rangle. \quad [\text{S2}]$$

Here the summation runs over all phonon branches $n$, $|k\rangle = e^{ikm}$ is a plane wave with the wave vector $k$. A scalar product is defined in the usual way, $\langle k|x_{nm}\rangle = \sum_m e^{-ikm} x_{nm}$. And we use a finite-width approximation to a δ-function to describe a finite lifetime. The result looks similar to the phonon dispersion of a periodic crystal, but the translation symmetry breaking due to the DW is manifested in the normal modes at a given frequency not having a single plane wave form with fixed $k$, but instead a combination of plane waves with different wave vectors.

The spectral function projected on the oscillations of $Q$, $\phi$, and $P$ modes resembles the bulk phonon dispersion with the dispersive branches for $\Gamma_2^-$ and $K_3$ phonons, along with several non-dispersive branches, corresponding to the phonons localized at the DWs. Here a damping parameter of 20 GHz, consistent with the optical experiments (*49*), is used in the calculations. The ripples in the false-color maps (Fig. S8A) are due to the finite size of the supercell. Fig. S8B shows the real-space oscillation of δ$Q$, δ$\phi$, and δ$P$ at several characteristic frequencies. The lowest branch, appearing in the GHz range, corresponds to the oscillations of the DW position around the equilibrium position, while the localized modes at the THz range correspond to the width oscillations of the $Q$, $\phi$, and $P$ textures. Note that DW-sliding mode is localized only perpendicular to the DW plane and free to propagate within the DWs. Our 1D calculation, therefore, only shows the bottom of the phonon band. The sliding mode with non-zero $k_y$, $k_z$ wave vectors in the DW plane are analogous to the acoustic waves in elastic media. The inter-site mode coupling terms similar to those with $c_1$, $c_2$ in Equation [S1] give rise to the phonon dispersion for the wave vector components in the DW plane (e.g. $yz$). These terms give rise to the band of DW-localized phonons, with phonon amplitudes in the DW plane having the oscillating form $\exp(k_y y + k_z z)$, and the bandwidth determined by the respective stiffness constants, and therefore of the same order as the bandwidth of $\Gamma_2^-$ and $K_3$ bands in the bulk.

The phonon calculations are performed using frozen-phonon method as implemented in Phonopy software, within the 3×3×2 supercell of the P6$_3$cm low-temperature unit cell using generalized gradient approximation to density functional theory (GGA, DFT) with a plane wave basis set and projector-augmented waves formalism as implemented in the Quantum Espresso package (*55, 56*). The plane wave cut-off of 35 Ry and density cut-off of 300 Ry are used. The magnetic ordering is approximated by A-type antiferromagnetism, which should be sufficient for the present estimates and LDA+U with atomic projection and U = 4 eV was used (*57*). The results of the phonon dispersion (Fig. S8C) and contributions from the $Q$, $\phi$, and $P$ modes are shown in Fig. S8D. The discrete model is then constructed using the force constants calculated, with the energy of the domain wall minimized in the harmonic approximation, which is justified by the small tilts of bi-pyramids from the equilibrium positions in the two domains. Then the energy is expanded around this state and the phonons are calculated. The results, projected on the phase and amplitude modes of trimerization and on polarization are shown in Fig. S8A.

This approach, combining models and first-principles calculations, allows us to treat the bulk and DW-specific phonons within the unified picture and clarifies the physics of the localized low-energy excitation in the SIM data.